\documentclass[amsmath,amssymb,amsbsy,reprint,prb,preprintnumbers,showpacs,superscriptaddress]{revtex4-2}
\usepackage{graphicx,color}
\usepackage{dcolumn}
\usepackage{bm}
\usepackage{braket}
\usepackage{mathtools}
\usepackage{ulem}
\usepackage[breaklinks,colorlinks=true,linkcolor=blue,urlcolor=blue,citecolor=blue]{hyperref}
\usepackage{times}
\usepackage{physics}
\usepackage{latexsym}
\usepackage{amsmath, amssymb}
\usepackage{mathtools}
\usepackage{multirow}

\newcommand{\be}{\begin{eqnarray}}
\newcommand{\ee}{\end{eqnarray}}

\begin{document}

\title{Intrinsic mixed state topological order in a stabilizer system under stochastic decoherence: Strong-to-weak spontaneous symmetry breaking from percolation point of view}
\date{\today}
\author{Yoshihito Kuno} 
\affiliation{Graduate School of Engineering science, Akita University, Akita 010-8502, Japan}
\author{Takahiro Orito}
\affiliation{Institute for Solid State Physics, The University of Tokyo, Kashiwa, Chiba, 277-8581, Japan}
\author{Ikuo Ichinose} 
\thanks{A professor emeritus}
\affiliation{Department of Applied Physics, Nagoya Institute of Technology, Nagoya, 466-8555, Japan}


\begin{abstract} 
Discovering quantum orders in mixed many-body systems is an ongoing issue. 
Very recently, the notion of an intrinsic mixed state topologically-ordered (IMTO) state was proposed. 
As a concrete example, we observe the emergence of IMTO by studying the toric code system under stochastic maximal decoherence by $ZX$-diagonal type projective measurement without monitoring. 
We study how the toric code state changes to an IMTO state at the level of the averaged quantum trajectories. 
This phase transition is understood from the viewpoint of spontaneous symmetry breaking (SSB) 
of 1-form weak symmetry, that is, the IMTO is characterized by the symmetry restoration from the SSB, 
which comes from the proliferation of anyons. 
To understand the emergent IMTO, order and disorder parameters of 1-form symmetry are numerically studied by stabilizer simulation.
The present study clarifies the existence of two distinct microscopic string operators for the fermionic anyons, that leads to distinct fermionic strong and weak 1-form symmetries, and also
the obtained critical exponents indicate strong relation between IMTO and percolation.
\end{abstract}


\maketitle
\section{Introduction}
Quantum devices such as quantum computers and quantum memory are significantly affected by quantum noise from external environment \cite{gardiner2000,preskill2018,dennis2002,wang2003,ohno2004}. However, under noise, an intermediate scale quantum device and computer \cite{ebadi2021,bluvstein2024} are expected to give some great ability beyond the ability of the classical ones \cite{preskill2018}.
Topologically ordered state \cite{wen2007,Wen_2017,Wen_text}, which is one of the non-trivial states of matter found first in condensed matter physics, 
gives a basic platform for topological quantum computations \cite{nayak} or quantum memories \cite{kitaev2003}. 
Topological order has been already realized in experimental quantum systems \cite{satzinger2021,acharya2022,zhao2022}. 
Decoherence and noise from environment are inevitable and change quantum states into undesired ones. 
However, decoherence can lead to rich physical phenomena, leading to an unconventional quantum state of matter beyond the pure state. In particular, 
decoherence can create some unconventional topologically-ordered mixed states in quantum many-body systems. 
Recently, as an intriguing development, the notion of intrinsic mixed state topological order (IMTO) was proposed, i.e., 
the mixed state possesses some topological order with no counterparts in any pure states \cite{wang2024,sohal2024,sang2024,chen2024}. 
Interestingly, the IMTO can be distinguished or classified from the aspects of higher-form symmetries \cite{gaiotto2015,mcgreevy2023} 
and their strong and/or weak spontaneous symmetry breaking (SSB). 
While the detailed notion of it has been discussed and its theoretical concept \cite{lee2023,lessa2024,sala2024,KOI2024,wang2024,sohal2024,liu2024_SSSB,guo2024,shah2024} 
is being developed, there are only a few concrete examples of emergent IMTO states studied by the numerical methods. 
We fill this gap in this letter. 

We shall give a detailed study on the emergent IMTO under a stochastic ZX-type decoherence from 
the two-dimensional (2D) toric code (TC) stabilizer state, the system of which can be efficiently simulated in the stabilizer formalism. 
Further, we numerically investigate the critical properties of the phase transition as increasing the spatial rate of the decoherence, by observing density matrix trajectories \cite{gullans2020} created through the large-scale stabilizer simulation \cite{gottesman1998,aaronson2004}. 
This phase transition is understood as a restoration from SSB
of 1-form weak symmetries, that is, the IMTO is understood as a {\it disordered phase} with respect to the 1-form weak symmetries, whereas the parent TC state 
is a strong-to-trivial SSB state in the sense that the magnetic and electric 1-form symmetries are all spontaneously broken
in both strong and weak symmetry senses~\cite{zhang2024}. 
As a result of the phase transition, the content of relevant anyons is drastically changed, and anyon proliferation
by the decoherence can be clearly understood from the perspective of the 1-form symmetries.
The numerical simulation corroborates our consideration on the IMTO, 
and furthermore, it indicates that the transition to the IMTO is understood as a `percolation of decoherence', supported by scaling analysis. 
Throughout this work, we study the IMTO. 
In literature, there are a few different definitions of `IMTO', and to classify the IMTO from various aspects is just an on-going issue. 
In this work, we mostly consider a decohered state with fermionic-anyon proliferation and call it an IMTO state. 
Thus, the IMTO used in this study is close to that of the recent work \cite{wang2024}.

The rest of this paper is organized as follows. 
In Sec.~II, we introduce the model under decoherence, described by a quantum channel operation \cite{Nielsen2011}. 
We start with the TC state and operate the $ZX$-type local decoherence.
In order to get an intuitive picture, we consider a mixed state under maximal $ZX$-type decoherence, and discuss 1-form-symmetry properties of states and channel. 
In Sec.~III, we introduce and explain target physical quantities to characterize the properties of mixed states under the decoherence.
To investigate decoherence effects and the evolution of mixed state from the TC state, we employ the quantum-trajectory-ensemble scheme  
with stochastic local decoherence, recently used in Refs.~\cite{gullans2020,weinstein2022}.
In Sec.~V, we explain the numerical protocol by using the efficient stabilizer algorithm. 
Through the numerical investigation, we find a clear decoherence-induced phase transition, in which the TC state evolves into an ensemble-averaged IMTO, first found in Ref.~\cite{wang2024}. 
We also study its criticality by using finite-size scaling analysis, etc.
There,  SSB of 1-form symmetries of the states plays an important role. 
In Sec.~VI, we discuss the numerical results obtained in Sec.~V for the viewpoint of percolation.
In Sec.~VII, we summarize the aspect of the 1-form symmetries of the  phase transition.
Section VIII is devoted to discussion and conclusion.

\section{Toric code system under ZX-decoherence}
\begin{figure}[t]
\begin{center} 
\vspace{0.5cm}
\includegraphics[width=8.8cm]{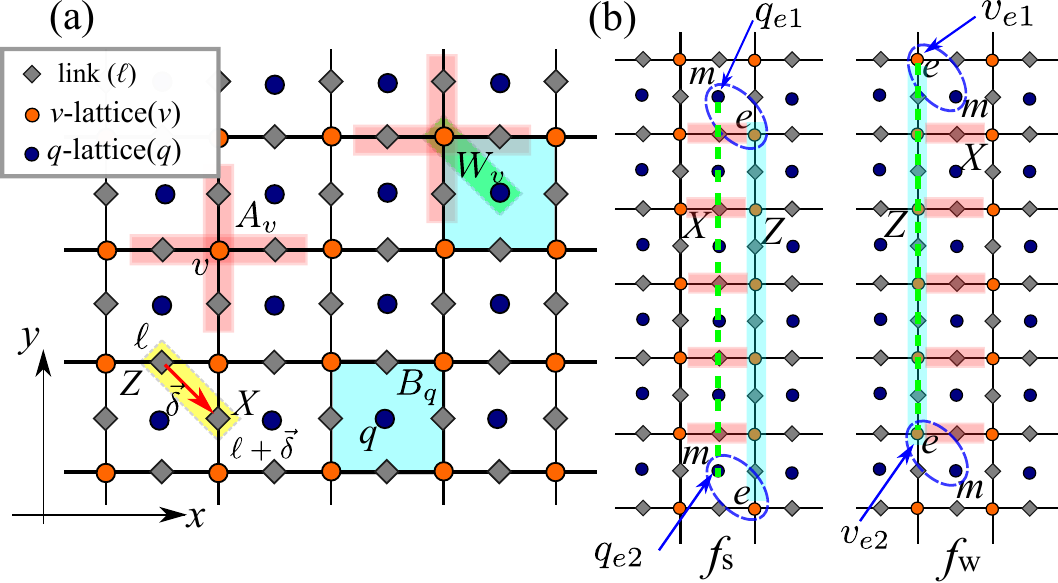}  
\end{center} 
\caption{(a) Schematic of toric code system. The red arrow is shift vector $\vec{\delta}$ used for definition of ZX and XZ operators. (b) The endpoints of open strings $\prod X_{\ell}Z_{\ell+\vec{\delta}}$  and $\prod Z_{\ell}X_{\ell+\vec{\delta}}$  create $f_s$ and $f_w$-anyons, respectively, both of which are a fusion of $e$ and $m$-anyons. Note that locations of $e$ and $m$-anyons are different in $f_s$ and $f_w$-anyons, that is, there are
two kinds of $f$-anyons.}
\label{Fig1}
\end{figure}

We consider the 2D TC system resided on torus. The system is shown in Fig.~\ref{Fig1}(a), where we introduce a lattice composed of $L_x \times L_y$ plaquettes ($q$-lattice) and $L_x \times L_y$ vertices ($v$-lattice) on the torus. 
Physical qubits reside on links of the $v$-lattice and then, the total number of qubit is $L\equiv 2L_xL_y$. 
We represent links by $\{\ell\}$, and 
we also use the link vector $\ell$ to the link number, that is, $\ell$ takes $\ell=0,1,2,\cdots, 2L_xL_y-1$. 

The initial state is set by the following stabilizer generators \cite{Nielsen2011},  
$S_{\rm int}=\{A_v | v\in \mbox{all vertex but} \ v_0 \} + \{ B_q|q\in 
\mbox{all plaquette but} \ q_0 \}$. 
This state is the maximally-mixed  ground state with the four-fold degeneracy of the TC on torus \cite{kitaev2003}. 
$A_v$ and $B_q$ are the star and plaquette operators; $A_v=\prod_{\ell_{v} \in v} X_{\ell_{v}}$ and $B_q=\prod_{\ell_{q} \in q} Z_{\ell_{q}}$, with $\ell_{v} \in v$ standing for links emanating from vertex $v$, and $\ell_{q} \in q$ for links composing plaquette $q$. 

For later calculations, we add the typical logical operator into the set of the stabilizer generator as
$\{S_{X-{\rm lg}} \}=\{\prod_{\ell\in {\gamma_{x}}}X_{\ell},\prod_{\ell\in {\gamma_{y}}}X_{\ell}\}$, where $\gamma_{x(y)}$ is $x$($y$)-directed non-contractible loop (the generator of 1-form electric $Z_2$ symmetry ('t Hooft loop)) on the $q$-lattice. The stabilizer state (pure state) $S_{\rm int}+S_{X-{\rm lg}}$ is a TC state. We denote the density matrix by $\rho_{\rm TC}$. 

For the 2D system, we consider the channel of a local maximal ZX-coherence given by 
\begin{eqnarray}
\mathcal{E}^{ZX}_\ell[\rho]=\sum_{\beta_\ell=\pm}P^{ZX}_{\beta_\ell}\rho P^{ZX\dagger}_{\beta_\ell}=\frac{1}{2}\rho+\frac{1}{2}Z_\ell X_{\ell+\vec{\delta}}\rho Z_{\ell}X_{\ell+\vec{\delta}}, \nonumber
\end{eqnarray}
where $\rho$ is a state (density matrix), $\vec{\delta}=(1/2,-1/2)$ and $\beta_i$ is a measurement outcome taking the value $\pm 1$. 
This decoherence corresponds to the projective measurement by $Z_{\ell}X_{\ell+\vec{\delta}}$ without monitoring as shown in Fig.~\ref{Fig1}(a). 
The operation $Z_{\ell}X_{\ell+\vec{\delta}}$ also creates a pair of $f$-anyons (dyon) on the TC state as we explain later on \cite{arakawa2004}. 

Let us first consider a maximal ZX-decohered state, that is, for all links, the ZX-decoherence channel is applied as 
$
\mathcal{E}^{all}[\rho]=\prod_{\ell} \mathcal{E}^{ZX}_\ell[\rho]
$. 
Then, the obtained state is $\rho_f\equiv \mathcal{E}^{all}[\rho_{\rm TC}]$. The state $\rho_f$ is readily described by the stabilizer formalism as $\rho_f\propto \prod^{L_x Ly-2}_{v=0}\frac{1+W_v}{2}$, where $W_v=A_{v}B_{v+\vec{\delta}}$ as shown in Fig.~\ref{Fig1}(a). 

In order to understand the mixed state $\rho_f$, we introduce two kinds of 1-form symmetry, symmetry operators of which are defined as follows $W^{ZX}_{\gamma_c}\equiv \prod_{\ell\in \gamma_c} Z_\ell X_{\ell+\vec{\delta}}$  and $W^{XZ}_{\gamma_c}\equiv \prod_{\ell\in \gamma_c} X_\ell Z_{\ell+\vec{\delta}}$  with 
a contractible loop $\gamma_c$ on the links of the $v$ lattice.
It is easily seen that the operator $W^{ZX}_{\gamma_c}$ is a product of consecutive Kraus operators,
 $\{Z_\ell X_{\ell+\vec{\delta}}\}$, and satisfies the 1-form weak-symmetry condition \cite{groot2022} 
 [see the definition of weak and strong symmetries in Appendix A.], $W^{ZX}_{\gamma_c}\rho_f W^{ZX}_{\gamma_c}=\rho_f$, although it {\it doe not} commutes some of Kraus operators.
On the other hand, the operator $W^{XZ}_{\gamma_c}$ commutes all of the Kraus operators and $W^{ZX}_{\gamma_c}$,
and satisfies the 1-form strong-symmetry condition \cite{groot2022},  $W^{XZ}_{\gamma_c}\rho_f=\rho_f W^{XZ}_{\gamma_c}=\rho_f$.
By employing open string ${\cal C}$ instead of the close loop $\gamma_c$ for the above operators, corresponding anyons 
emerge at the endpoints of ${\cal C}$ as shown in Fig.~\ref{Fig1}(b).
We call them $f_w$ and $f_s$, respectively.
We remark that both the anyons are obtained by the fusion of $e$-anyon and $m$-anyons but their geometric location is different in $f_w$ and $f_s$. 
This point has been overlooked so far and plays an important role in understanding $\rho_f$. 
It is not so difficult to show $W^{ZX}_{\cal C}\rho_f  W^{ZX}_{\cal C}=\rho_f$, which implies emergence of $f_w$-anyon proliferation
in $\rho_f$~\cite{wang2024,sohal2024}.
On the other hand, the operator $W^{XZ}_{\cal C}$ does not commute with the stabilizer $W_v$ at endpoints, and therefore, \textit{$f_s$-anyon is a detectable anyon behaving as a stable excited state}, 
whereas \textit{$f_w$-anyon is an undetectable anyon to proliferate} \cite{ellison2023}. 
The states $\rho_{\rm TC}$ and $\rho_f$ are classified from the viewpoint of SSB/non-SSB of the 1-form symmetries proposed recently \cite{zhang2024}. 
The 1-form symmetry aspect is discussed in detail in Appendix B,
and it is emphasized there that the order and disorder parameters of the 1-form symmetries give us useful measures for the phase transition from the TC state (topological SSB state) to the IMTO state.

Study on the 1-form symmetry is still an on-going subject.
Therefore, before going into details of the investigation, we remark on the 1-form symmetry and status of the present work. 
In some cases, 1-form symmetries emerge in an `low-energy' effective model even if its original system does not possess corresponding symmetry~\cite{mcgreevy2023}. 
Furthermore,  such a 1-form symmetry is sometimes explicitly or spontaneously broken, whereas it induces a symmetry order such as SPT order, etc~\cite{mcgreevy2023,Verresen2024}. 
Throughout this study, we do not consider `effective' 1-form symmetries, instead, we focus on the system, in which 1-form symmetries explicitly appears, and we study its strong or weak symmetry properties~\cite{groot2022}.

\section{Target physical quantities}
To understand the effect of the ZX-decoherence and detect the IMTO, we focus on analyzing the following physical quantities. \\

\subsection{Specific behavior of entanglement negativity}
We firstly focus on the entanglement negativity for a contractible closed subsystem given as 
$N_A\equiv \log_2|\rho^{\Gamma_A}|_1$ \cite{peres1996,Horodecki_family}, where $|\cdot|_1$ is the trace norm and $\Gamma_A$ is a partial transpose for A-subsystem. 
The negativity can reveal a transition of states under decoherence by quantifying quantum correlations in mixed states \cite{lu2020,sang2021,weinstein2022}.
The practical method of numerical calculation in the stabilizer formalism is explained in Appendix C 2.
As the A-subsystem, we use a rectangular subsystem on the $v$-lattice of size $\ell_x \times \ell_y$, with $\ell_x=2k_A$ and $\ell_y=2$ ($k_A\in \mathbb{N}^0$) as in Fig.~\ref{Fig_NA}(b).
Then as in Ref.~\cite{wang2024}, a specific behavior of the negativity is expected for the maximal $ZX$-decoherence limit $\rho_f$, i.e., the subsystem size scaling of the negativity given by  $N^f_A(k_A)=\frac{N_P}{2}-\frac{1}{2}$ with $N_P=6(k_A+1)$. 
We shall observe whether such a scaling holds in the $ZX$-decohered system later.
Furthermore, as the negativity is a measure of the long-range entanglement, its comparison with the behavior of the logical operator is interesting. 
We shall discuss this point after showing the numerical results.\\

\subsection{Order and disorder parameters for 1-form weak SSB} 
To characterize the IMTO, we shall make use of a 1-form symmetry picture and its SSB, the generator of which is given by $W^{ZX}_{\gamma_{c}}$, here we only consider a contractible loop $\gamma_{c}$ on the $v$-lattice as before.

The system density matrix $\rho$ starting with $\rho_{\rm TC}$ keeps the 1-form weak symmetry 
$W^{ZX}_{\gamma_c}\rho W^{ZX\dagger}_{\gamma_c}=\rho$ \cite{groot2022,ma2024} through the ZX-channels $\mathcal{E}^{ZX}_{\ell}$, since the symmetry generator 
$W^{ZX}_{\gamma_c}$ anti-commutes with some of Kraus operators but commutes with all of the stabilizers. 
Based on this symmetry, its SSB can be detected by the following two observables;
Order and disorder parameters, details of which are explained in Appendices.
The utility of these is to be examined as they have been proposed very recently~\cite{zhang2024}, 
and therefore, the present study works as an efficient example to observe their reliability.
Please see Appendix B.

As suggested by Ref.~\cite{zhang2024}, 
the order parameter of the 1-form weak SSB is given by 
\begin{eqnarray}
C^{\rm I}(\rho,\gamma'_c) \equiv \Tr[\rho W^{Z}_{\gamma'_c} \rho ]/\Tr[\rho^2],\nonumber
\end{eqnarray}
where $\gamma'_c$ is a contractible loop on the $v$-lattice, and $W^{Z}_{\gamma'_c}\equiv \prod_{\ell\in \gamma'_c}Z_\ell$ is a Wilson loop, 
one of conjugate operators to $W^{ZX}_{\gamma_c}$, coming from the fact that $e$-anyon and $f_w$-anyon
have nontrivial braiding relation.

Here, the value of $C^{\rm I}(\rho,\gamma'_c)$ is determined such that if $W^{Z}_{\gamma'_c}$ is an element of stabilizer of $\rho$, then $C^{\rm I}(\rho,\gamma'_c)=1$ and for the density matrix with a local ZX-decoherence acting on the loop $\gamma'_c$, $C^{\rm I}(\rho,\gamma'_c)=0$.
The finite value of $C^{\rm I}(\rho,\gamma'_c)$ for a large loop $\gamma'_c$ means the 1-form weak SSB. 
In fact, the initial state $C^{\rm I}(\rho_{\rm TC},\gamma'_c)=1$.
For later numerics, we numerically calculate the average sum of $C^{\rm I}(\rho,\gamma'_c)$ for different typical square loop $\gamma'_c(k)$ (the side of the length is $\ell_x=\ell_y=k$), and define
$
\chi^{\rm I}[\rho]\equiv\frac{1}{N_{\ell}}\sum^{N_\ell}_{k=1}C^{\rm I}(\rho,\gamma'_c(k)) 
$ with $N_{\ell}=\min\{L_x-2,L_y-2\}$.

The disorder parameter is given by  
\begin{eqnarray}
C^{{\rm II}}(\rho,(v_{e1},v_{e2}))\equiv \frac{\Tr[\rho W^{ZX}(v_{e1},v_{e2}) \rho W^{ZX}(v_{e1},v_{e2})]}{\Tr[\rho^2]},
\nonumber
\end{eqnarray}
where $W^{ZX}(v_{e1},v_{e2})$ is an open string operator $W^{ZX}_{\cal C}$ for ${\cal C}$ of end points $v_{e1}$ and $v_{e2}$,
as shown in Fig.~\ref{Fig1}(b). 
The observable $C^{{\rm II}}(\rho,(v_{e1},v_{e2}))$ is an extension of the R\'enyi-2 correlator used for
study on weak symmetry SSB, first suggested in \cite{zhang2024}.
The previous type of the R\'enyi-2 correlator is that (I) the operator is a local charged operator for some global (i.e., 0-form) symmetry, and any loop operators  have not been considered, and (II) so far, the R\'enyi-2 correlator has been used to identify the strong-to-weak symmetry breaking in `spin' systems~\cite{lee2023, lessa2024,sala2024,KOI2024}. 
For the initial state $C^{{\rm II}}(\rho_{\rm TC},(v_{e1},v_{e2}))=0$, which means that TC state is strong-to-trivial SSB \cite{zhang2024}. 
We shall calculate this quantity under the ZX decoherence channel to observe a transition to the IMTO. 
For later numerics, we numerically calculate the susceptibility defined by
$
\chi^{\rm II}(\rho)\equiv\frac{1}{L_x (L_y-3)}\sum^{L_x-1}_{i_x=0}\sum^{{L_y-3}}_{\ell=1}C^{{\rm II}}(\rho,((i_x,0),(i_x,\ell))) 
$
for a state $\rho$, where $(i_x, 0)$ and $(i_x,\ell)$ are the locations of $e$-anyons residing on the endpoints of the open string.

\section{Stochastic ZX-decoherence-only stabilizer channel and trajectory ensemble}
We consider a stabilizer channel consisting of solely ZX-decoherences, where the local ZX-decoherence at diagonal links $\mathcal{E}_\ell$ is applied with a probability $r$ ($0\leq r \leq 1$) for all links $\ell$, that is, we consider a single-layer decoherence. 
Such local decoherence is numerically tractable in the algorithm even for large system sizes \cite{weinstein2022}, as explained in Appendix C 1. 
In the system under the decoherence, we record locations of the occurrence of decoherence but no other information of the channel. 
The system with this protocol exhibits a single trajectory of the state labeled by $s$, described as
$
\rho^s_{D}=\mathcal{E}^{ZX}_{\ell_0}\circ 
\mathcal{E}^{ZX}_{\ell_1}\circ \cdots \circ \mathcal{E}^{ZX}_{\ell_{N_D}}[\rho_0]$, 
where $\rho^s_{D}$ is the finial mixed state after the single-layer decoherence, $\rho_0$ is an initial state, $\mathcal{E}^{ZX}_{\ell_k}$ is the ZX decoherence at a position $\ell_k$, and $N_D$ is the number of $\mathcal{E}^{ZX}_{\ell_k}$ performed with the probability $r$ ($N_D \sim r \times [\mbox{ total number of site}]$). 
Similar setups were used in Refs.~\cite{gullans2020,weinstein2022,liu2024,liu2024v2,KOI2024}. 
Here, we note that 
the channel using the trajectory-sampling scheme is different from the channel considered in the previous studies \cite{Fan2024,wang2024}, where there is no notion of sampling, that is, prescription without recording outcomes where to apply the local decoherence.
This difference leads to consequence that some physical quantities obtained by averaging over the samples of the trajectory density matrix $\rho^s_D$  might not coincide with those obtained by the density matrix applied by the channel without recording the position of the decoherence, considered in the previous studies \cite{Fan2024,wang2024}. 
This point has already been commented on in Ref.~\cite{KOI2024}. 
This trajectory treatment that we choose is the same as the approach in Refs.~\cite{gullans2020,weinstein2022,liu2024,liu2024v2,KOI2024}. 
In particular, we expect that physical quantities obtained by tracing the nonlinear quantities of the density matrix (just as $C^{\rm I}(\rho,\gamma'_c)$ and $C^{{\rm II}}(\rho_{\rm TC},(v_{e1},v_{e2}))$) might exhibit a different value for the transition probability and its criticality, etc, \cite{KOI2024} compared to those obtained by the density matrix employed in \cite{Fan2024,wang2024}.

By using these trajectory samples $\{\rho^s_{D}\}$, we obtain the ensemble average of the physical quantities denoted as $Q(\rho^s_D)$ introduced in the previous sections.
This quantity is given by the non-linear form of the density matrix \cite{non_linear_Q}, denoted by $\mathbb{E}[Q(\rho^s_D)]$,
where $\mathbb{E}[\cdot]$ means averaging over the samples of the trajectory density matrix $\rho^s_D$.
We shall show that the IMTO state appears in the level of quantum trajectory ensemble.

\begin{figure}[t]
\begin{center} 
\vspace{0.5cm}
\includegraphics[width=9cm]{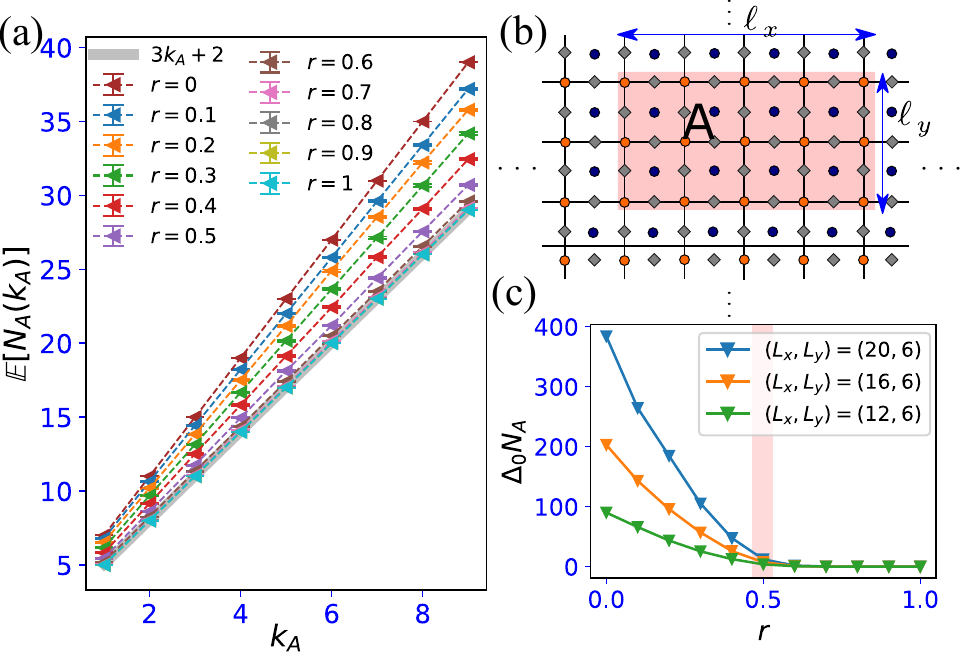}  
\end{center} 
\caption{(a) $\mathbb{E}[N_A(k_A,r)]$ for different $r$'s. (b) The subsystem partition for $N_A$. The rectangle edges are $\ell_x=2k_A$ and $\ell_y=2$ with $k_A\in \mathbb{N}^0$ 
(c) The behavior of $\Delta_0 N_A(r,L_x)$. The number of sample trajectories is $\mathcal{O}(10^2)$.}
\label{Fig_NA}
\end{figure}

\section{Numerical results for averaged physical quantities} 
We shall show the numerical results in the stochastic ZX-decoherence circuit. \\

\subsection{Averaged entanglement negativity} 
We plot subsystem-size dependence of $\mathbb{E}[N_A(k_A)]$ on $k_A$ for various $r$'s with fixing the system size as $(L_x,L_y)=(20,6)$. 
As shown in Fig.~\ref{Fig_NA}(a), we first observe that $\mathbb{E} [N_A(k_A)]$ shows monotonic decrease as increasing $r$. 
The reason for this decreasing behavior of $\mathbb{E}[N_A(k_A)]$ is explained in Ref. ~\cite{wang2024}, i.e., the contribution to negativity is determined by the number of cuts in edges of stabilizers by the subsystem boundary;
If the $W_v$ stabilizers are formed around the boundary of the subsystem, the number of cuts for the $W_v$ edges determines the negativity. 
Here, the number of subsystem cuts for each $W_v$ stabilizer edge in the state $\rho_f$ is smaller than the number of cuts for the stabilizer edges of $A_v$ and $B_q$ in the TC phase. Thus negativity is reduced if the stabilizer $W_v$ is proliferated.

Furthermore, the $k_A$ linear scaling line of $\mathbb{E}[N_A]$ rapidly approaches the scaling line of the maximal $ZX$-decoherence state $\rho_f$ \cite{wang2024}. 
This rapid approach is a signature of the phase transition to the IMTO having the same long-range entanglement (LRE) with $\rho_f$. 
Location of the critical point is elucidated by observing the difference between the analytical calculation
$N^f_A$ and numerically-obtained one $\mathbb{E}[N_A(k_A)]$ defined as
$\Delta_0 N_A(r)\equiv \sum^{(L_x-2)/2}_{k_A=1}(\mathbb{E}[N_A(k_A,r)]-N^f_A)^2$. 
The data is displayed in Fig.~\ref{Fig_NA}(c), which shows $\Delta_0 N_A( r=0.5)\approx 0$ for the various system sizes, and 
we estimate the critical point as $r_c=0.5$.

The system-size independent term in $N^f_A$ indicates the existence of the LRE even for $r>r_c$.
We also investigated the behavior of the logical qubit, i.e., the annihilation probability of the logical operators of the TC under the ZX-decoherence, and found that it disappears at $r\sim 0.5$ (See Appendix D).
As suggested in \cite{wang2024}, this is a typical behavior of the IMTO, and the origin of the LRE is to be clarified in a later research. 
We also comment that this rapid approach observed in $\Delta_0 N_A(r)$ can originate from the ensemble-averaged quantity obtained from the stochastic trajectory samples. 
For example, let us consider an extreme example, that is, the measurement-induced entanglement phase transition (MIPT)~\cite{Fisher_2023}. 
There, the MIPT is not observed by an entanglement entropy (EE) calculated using the averaged density matrix. 
On the other hand, if we observe the ensemble average of the EE obtained by each density matrix trajectory, we can recognize the phase transition.

\subsection{Averaged 1-form weak SSB restoring transition} 
Let us study how the SSB properties of the system change under the decoherence.
For the order parameter $C^{\rm I}(\rho,\gamma'_c)$, the initial state $\rho_{\rm TC}$ has $C^{\rm I}(\rho_{\rm TC},\gamma'_c)=1$ since the closed loop $W^Z_{\gamma'_c}$ is 
given by a product of $B_p$'s and is an element of the stabilizer group of the TC. 
This fact means that the state $\rho_{\rm TC}$ is a 1-form weak SSB state \cite{come_TC_SSB}.
By the decoherence $\mathcal{E}^{ZX}_\ell$ with Kraus operators, some of which anti-commute with the operator $W^Z_{\gamma'_c}$, $W^Z_{\gamma'_c}$ tends to disappear in the stabilizer group as $r$ is getting large, inducing $C^{\rm I}(\rho_{\rm TC},\gamma'_c)\to 0$. 
[The numerical calculation method of $C^{\rm I}$ is explained in Appendix C 3]. 

\begin{figure}[t]
\begin{center} 
\vspace{0.5cm}
\includegraphics[width=7.5cm]{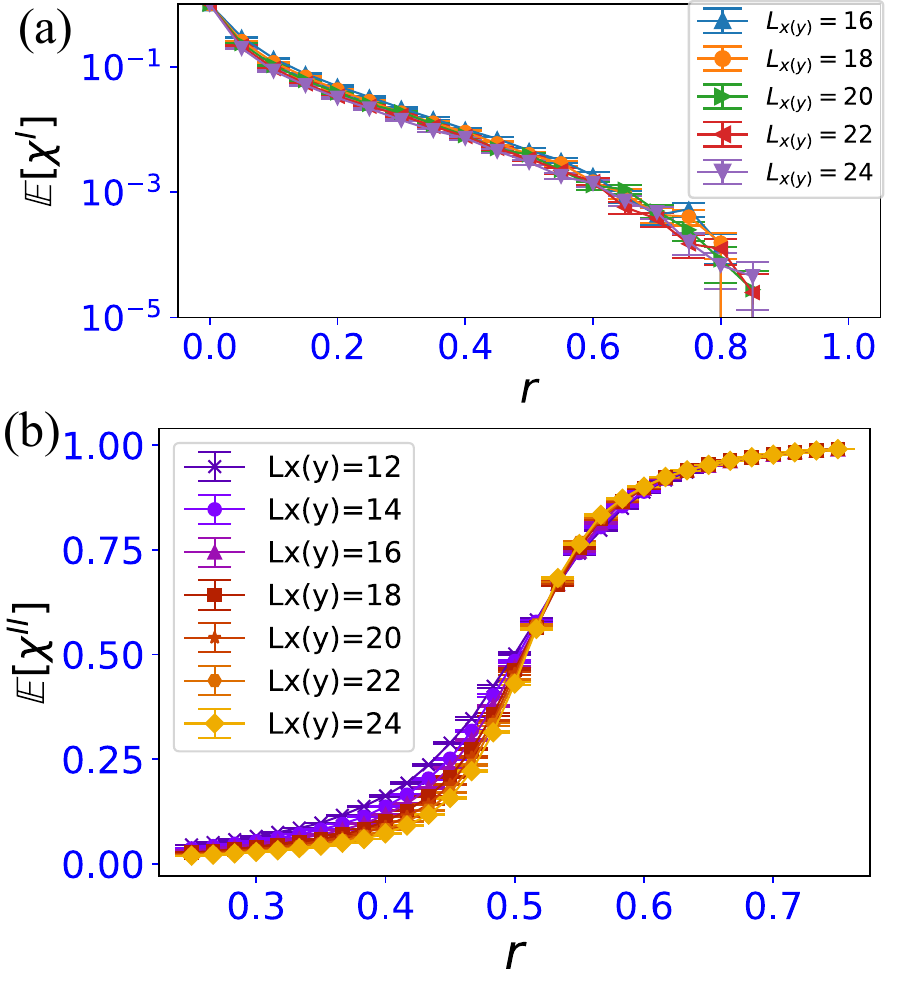}  
\end{center} 
\caption{(a) The behavior of $\mathbb{E}[\chi^I]$. (b) The behavior of $\mathbb{E}[\chi^{\rm II}]$. For each data, we averaged over $\mathcal{O}(10^3)$ samples.}
\label{Fig3}
\end{figure}

For the disorder parameter $C^{{\rm II}}(\rho,(v_{e1},v_{e2}))$, certain $A_v$ and $B_p$ stabilizers of $\rho_{\rm TC}$ anti-commute with both endpoints of the open string $W^{ZX}(v_{e1},v_{e2})$, resulting in $C^{{\rm II}}(\rho_{\rm TC},(v_{e1},v_{e2}))=0$ 
[The method of the calculation of the R\'enyi-2 correlator is shown in Appendix C 4.].
As increasing $r$, the operator $W_{v}$ proliferates and becomes a stabilizer generator of the mixed state. 
Then, the open string $W^{ZX}(v_{e1},v_{e2})$ tends to commute with the set of the stabilizer generators of the decohered state instead of $A_v$ and $B_p$. 
In the limit $r=1$, the state $\rho_f$ commutes with any open string $W^{ZX}(v_{e1},v_{e2})$, leading to $C^{{\rm II}}(\rho_f,(v_{e1},v_{e2}))=1$. 
Thus, the stochastic circuit tends to increase the value of $C^{{\rm II}}(\rho,(v_{e1},v_{e2}))$ and the decohered state turns into a state restoring the 1-form weak symmetry. 
Note that since the open string operator $W^{ZX}(v_{e1},v_{e2})$ creates $f_w$-anyon at the endpoints, the finite value of $C^{{\rm II}}(\rho_f,(v_{e1},v_{e2}))$ means that the $f_w$-anyons proliferate in the state through the decoherence.

\begin{figure}[t]
\begin{center} 
\vspace{0.5cm}
\includegraphics[width=8cm]{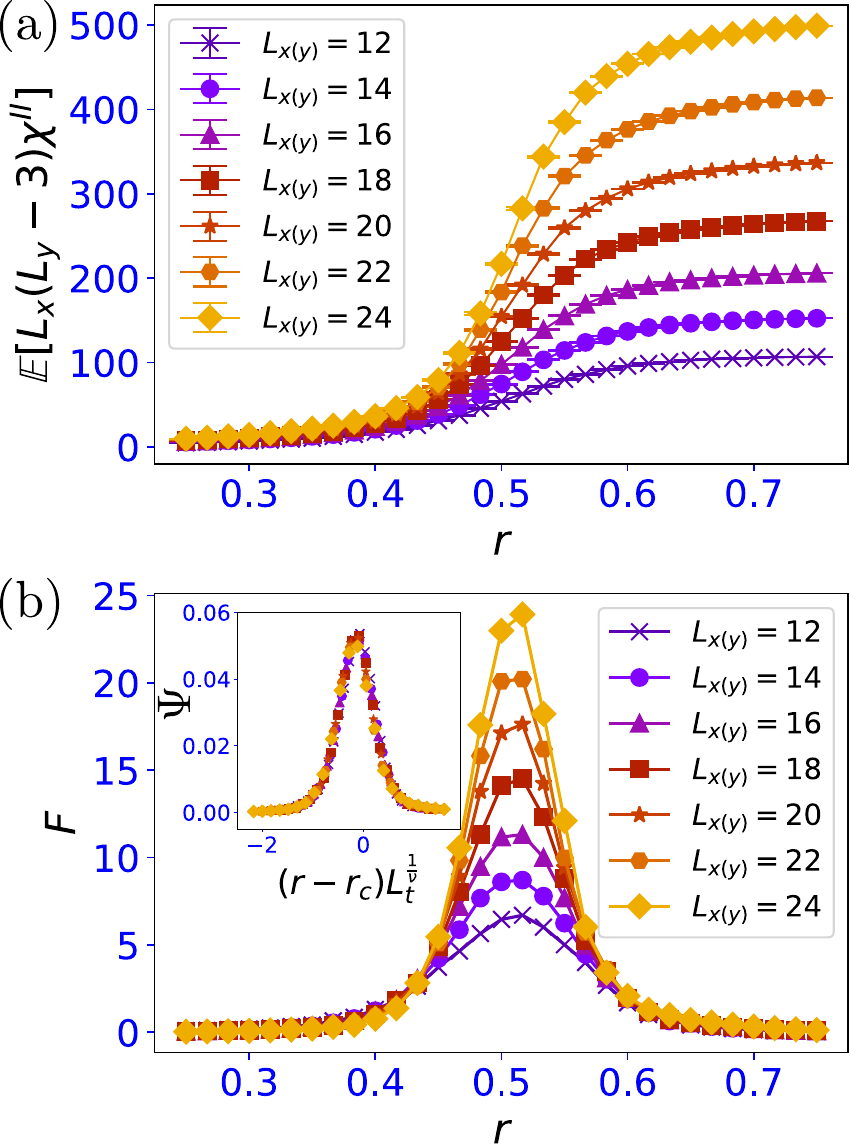}  
\end{center} 
\caption{(a) Behavior of $\mathbb{E}[L_x(L_y-3)\chi^{\rm II}]$ for various system sizes. 
(b) Rescaled variance $F$ for various system sizes. 
Inset: the scaling collapse for the rescaled variance $F$.
For each data, we averaged over $\mathcal{O}(10^3)$ samples.}
\label{Fig4_v2}
\end{figure}
Based on the above observation, let us numerically study the average of $\chi^I$ and $\chi^{\rm II}$, 
$\mathbb{E}[\chi^{\rm I}(\rho_s,r)]$ and $\mathbb{E}[\chi^{\rm II}(\rho_s,r)]$, to verify the behavior of the order and disorder parameters. 

As shown in Fig.~\ref{Fig3}(a), $\mathbb{E}[\chi^{\rm I}(\rho_s,r)]$ starts from the state $\rho_{\rm TC}$ with $\chi^I=1$ and decreases as increasing $r$, indicating the vanishing of the weak SSB order for the 1-form weak symmetry of the generator $W^{ZX}_{\gamma_c}$. 
However, $\mathbb{E}[\chi^{\rm I}(\rho_s,r)]$ does not exhibit any particular behaviors in the vicinity of the expected transition point $r\sim 0.5$.
This might be not so surprising as the genuine expectation value of the Wilson loop, Tr$( W^Z_{\gamma'_c}\rho)$, does not
show any indication of the decoherence transition \cite{bao2023}. 
Therefore, study using disorder parameter is quite useful for observing the decoherent transition.

In addition, we comment on the behavior of the ensemble average value of $\mathbb{E}[C^{\rm I}(\rho_{s},\gamma'_c)]$. As mentioned in Sec. III A, if $\rho_s$ has a local ZX-decoherence acting on the loop $\gamma'_c$, then $C^{\rm I}(\rho_{s},\gamma'_c)=0$. From this fact, as the perimeter of the loop $\gamma'_c$ is larger, $C^{\rm I}(\rho_{s},\gamma'_c)$ tends to vanish. In the ensemble average for the probability $r$, $\mathbb{E}[C^{\rm I}(\rho_{s},\gamma'_c)]$ is estimated by $\mathbb{E}[C^{\rm I}(\rho_{s},\gamma'_c)]=(1-r)^{\mbox{(perimeter of $\gamma'_c$)}}$. 
Thus, $\mathbb{E}[C^{\rm I}(\rho_{s},\gamma'_c)]$ with the large loop of $\gamma'_c$ exponentially decreases for a finite $r$. Thus, it fails to characterize some phase transition for any $r$.

As explained in the above, the quantity $\mathbb{E}[\chi^{\rm II}(\rho_s,r)]$ is more suitable for detecting the phase transition of the mixed state as shown in \cite{KOI2024}. 
The numerical data of $\mathbb{E}[\chi^{\rm II}(\rho_s,r)]$ is shown in Fig.~\ref{Fig3}(b). 
We find that the value of $\mathbb{E}[\chi^{\rm II}(\rho_s,r)]$ exhibits a sudden change around $r=0.5$, indicating the disordered state for the 1-form weak symmetry of the generator $W^{ZX}_{\gamma_c}$ appears as increasing $r$. 
We find that combining this behavior with the result of $N_A$ in Figs.~\ref{Fig_NA}(a) and  (c), 
the transition from the TC state to the IMTO state takes place at $r\simeq 0.5$. 

We furthermore plot the ensemble average of rescaled $\chi^{\rm II}$ given by $\mathbb{E}[L_x(L_y-3)\chi^{\rm II}]$ and its rescaled variance as varying $r$ in Fig.~\ref{Fig4_v2}(a). 
It increases as the system size increases in the whole parameter region.
Then in Fig.~\ref{Fig4_v2}(b), we show its variance $\sigma$ calculated for many samples of 
$(L_x(L_y-3)\chi^{\rm II}(\rho_s,r))$ with various system sizes [${\rm var}[L_x(L_y-3)\chi^{\rm II}]$]. 
We then observe a rescaled variance $F\equiv \sigma/(L_x(L_y-3))$.
The peaks for various system sizes are located around $r=0.5$ and the value of the peak gets larger as increasing the system size. 
Through the variance $\sigma$ of $(L_x(L_y-3)\chi^{\rm II}(\rho_s,r))$ for various system sizes, we estimate the critical probability of the phase transition $r_c$, and we also study its criticality. 
To this end, we employ the rescaled variance $F\equiv \sigma/(L_x(L_y-3))$ and the finite-size scaling ansatz given such as $F \equiv L_{t}^{\frac{\zeta}{\nu}}\Psi((r-r_c)L_{t}^{\frac{1}{\nu}})$,
where the system typical length $L_{t}$, and $\zeta$ and $\nu$ are critical exponents. 
In particular, we are interested in the exponent $\nu$ (correlation-length critical exponent). 
We set $L_t=L_x$ and employ pyfssa package \cite{melchert2009,pyfssa}.
The best scaling collapse data in our numerics is shown in the inset of Fig.~\ref{Fig4_v2}(b). 
We here estimate $r_c=0.528\pm 0.007$, $\nu=1.36\pm 0.47$ and $\zeta=2.65\pm 0.37$. 
We find that the exponent $\nu$ is very close to that of the 2D percolation $4/3$ \cite{stauffer2018}. 
Thus, we expect that the phase transition found here relates to 2D classical percolation. 
In fact in the previous paper \cite{KOI2024}, we discussed how the decoherence transition is related to the percolation
from the viewpoint of stabilizer channel, and that observation can be applied to the present `gauge system'.

\section{Percolation picture}
As we observed in Sec.~V, the numerical study of the R\'{e}nyi-2 correlator $C^{\rm II}(\rho,v_{e1},v_{e2})$ and the scaling analysis indicate close relationship between the decoherence phenomenon and the 2D percolation.
In fact in our previous work \cite{KOI2024}, in which the 2D Ising and cluster models under decoherence are considered, we explain percolation picture derived by careful look at the R\'{e}nyi-2 correlator.
There, we focused on the zero-form symmetries and their strong-to-weak SSB \cite{SWSSB}.
It is expected that this observation sheds new light on the decoherence phenomenon in quantum many-body systems.
Therefore, it is interesting to see how the percolation picture works on the 1-form symmetry in the present model.
There is a hint about that issue; the TC model under $X$-decoherence is dual to 2D Ising model applied by $ZZ$-decoherence, and a $X$ string in the TC defined on an open string ${\cal C}$ is dual to a pair of $Z$ spins in Ising model located at edges of ${\cal C}$.
However, the detailed study on the R\'{e}nyi-2 correlators of string operators has been lacking so far.

\begin{figure}[t]
\begin{center} 
\vspace{0.5cm}
\includegraphics[width=8.5cm]{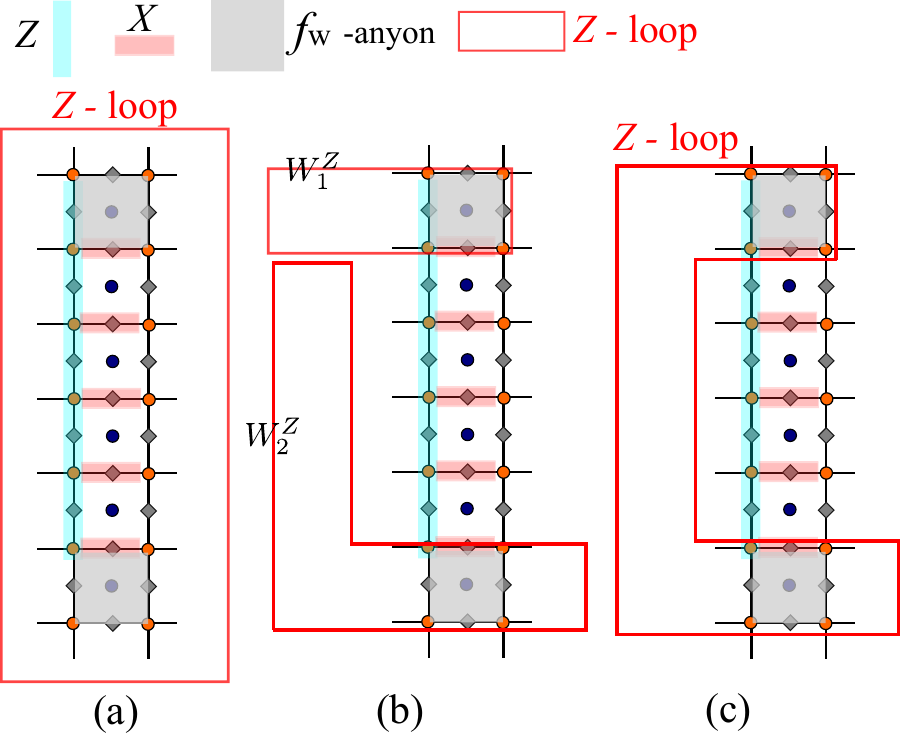}  
\end{center} 
\caption{Percolation picture: (a) String operator $W^{ZX}(v_{e1}, v_{e2})$ is totally located inside of  $Z$-loop stabilizer, and therefore, these two non-local operators commute with each other and a finite value of $C^{\rm II}$ appears. 
(b) Two $f_w$-anyons are encompassed by two different $Z$-loops and then, the string operator anti-commutes with each of $W^Z_1$ and $W^Z_2$, resulting $C^{\rm II}=0$. 
(c) Two $f_w$-anyons are encompassed by single $Z$-loop. 
Even though the middle portion of the string is not encompassed by $Z$-loop, two non-local operators commute with each other and a finite value of $C^{\rm II}$ appears. 
The above pictures eloquently show the relationship between 1-form symmetry under decoherence and percolation.}
\label{Fig5}
\end{figure}
In this section, we focus on the percolation picture of $C^{\rm II}(\rho,v_{e1},v_{e2})$ that is numerically studied in Sec.~V.
To this end, we employ the stabilizer generators such as 
$S'_{\rm int}=\{W_v | v\in \mbox{all vertex but} \ v_0 \} + \{ B_q|q\in 
\mbox{all plaquette but} \ q_0 \}$ to simplify the discussion, 
stabilizer group of which is obviously the same with that of $S_{\rm int}$. 
As explained in Sec.~III B, under $ZX$-decoherence, the set of stabilizers $\{B_q\}$ tend to merge with each other, and loop operators such as $\{\prod_{\ell\in \gamma} Z_{\ell}\}$ emerge, inside of which are void of $Z$-operators, whereas $\{W_v\}$ remain intact.
In that state, in order that $C^{\rm II}(\rho,v_{e1},v_{e2})$ would have a finite value, the open string operators $W^{ZX}(v_{e1}, v_{e2})$ has to commutes with all of the stabilizers of the state $\rho$.
The obstacle for this condition comes from the $Z$-loop stabilizers, and there are several possible geometrical cases to be considered.

In Fig.~\ref{Fig5}, we display three kinds of typical cases.
In case (a), the operator $W^{ZX}(v_{e1}, v_{e2})$ is totally located inside of the $Z$-loop, and therefore, these two operators obviously commute with each other giving a finite value of $C^{\rm II}$.
More precisely, the point is that two $f_w$-anyon operators at the edges of  $W^{ZX}(v_{e1}, v_{e2})$ are encompassed by the $Z$-loop.
In case (b), on the other hand, the two $f_w$-anyons are located inside of two different $Z$-loop stabilizers, $W^Z_1$ and $W^Z_2$, and therefore, $W^{ZX}(v_{e1}, v_{e2})$ anti-commute with both of them giving vanishing $C^{\rm II}$.
The third case (c) is the most important one for the present discussion,
$W^{ZX}(v_{e1}, v_{e2})$ is not fully located inside of a single $Z$-loop, but two $f_w$-anyons are encompassed by the loop.
In this case, $W^{ZX}(v_{e1}, v_{e2})$ commute with the $Z$-loop, and a finite value of $C^{\rm II}$ appears.
From the above discussion and figures in Fig.~\ref{Fig5}, it is obvious that the R\'{e}nyi-2 correlator of $W^{ZX}(v_{e1}, v_{e2})$ is closely related to 
the 2D site percolation problem on the dual lattice, i.e., $Z$-loop stabilizers correspond to generated percolation clusters, and percolation takes place if two largely separated sites $(v_{e1}, v_{e2})$ are encompassed in a single loop, inside of which corresponds to a cluster of percolation phenomenon.

In fact, the above pictorial discussion can be made more rigorous by the analytical methods.
The target measure is the R\'{e}nyi-2 correlator in Sec.~III B,
\begin{eqnarray}
&&C^{\rm II}({\cal E}^{ZX}(\rho_{\rm TC}), (v_{e1},v_{e2}))\nonumber\\
&&\propto \mbox{Tr} [{\cal E}^{ZX}(\rho_{\rm TC})W^{ZX}_{12}{\cal E}^{ZX}(\rho_{\rm TC})W^{ZX}_{12}],  
\nonumber
\end{eqnarray}
where we explicitly show the decoherence operator and use the abbreviation $W^{ZX}_{12}$ for $W^{ZX}(v_{e1},v_{e2})$.
For any Pauli strings $P^2=1$ and density matrix $\rho$, it is easily verified that $P(\rho+P\rho P)P=\rho+P\rho P$, and therefore,
$Z_\ell X_{\ell+\vec{\delta}} {\cal E}^{ZX}_\ell (\rho) Z_\ell X_{\ell+\vec{\delta}}={\cal E}^{ZX}_\ell (Z_\ell X_{\ell+\vec{\delta}}\rho Z_\ell X_{\ell+\vec{\delta}})
={\cal E}^{ZX}_\ell (\rho)$.

Now, let us consider the the straight string operator $W^{ZX}_{12}$ in Fig.~\ref{Fig5}.
For any string operator $\tilde{W}^{ZX}_{12}$ connecting two end points $(v_{e1}, v_{e2})$ without a breaking, 
$W^{ZX}_{12} \rho_{\rm TC} W^{ZX}_{12}= \tilde{W}^{ZX}_{12} \rho_{\rm TC} \tilde{W}^{ZX}_{12}$,
as the combined closed loop operator $(W^{ZX}_{12}\tilde{W}^{ZX}_{12})$ is an element of the stabilizer group of the toric code state $\rho_{\rm TC}$. 
There are two kinds of cases to be considered separately as the above pictorial discussion indicates.
In the first one, ${\cal E}^{ZX}(\cdot)$ totally contains the string operator $\tilde{W}^{ZX}(1,2)$.
In this case, the correlator is finite as the following equation shows,
\begin{eqnarray}
&&C^{\rm II}({\cal E}^{ZX}(\rho_{\rm TC}), (v_{e1},v_{e2}))\nonumber\\
&&\propto \mbox{Tr} [{\cal E}^{ZX}(\rho_{\rm TC}) {\cal E}^{ZX} (W^{ZX}_{12} \rho_{\rm TC}W^{ZX}_{12})]
\nonumber \\
&&=\mbox{Tr} [{\cal E}^{ZX}(\rho_{\rm TC}) {\cal E}^{ZX} (\tilde{W}^{ZX}_{12} \rho_{\rm TC}\tilde{W}^{ZX}_{12})]  \nonumber \\
&&= \mbox{Tr} [{\cal E}^{ZX}(\rho_{\rm TC}) {\cal E}^{ZX}(\rho_{\rm TC})].
\nonumber
\end{eqnarray}
With proper normalization, it is straightforward to show that $C^{\rm II}({\cal E}^{ZX}(\rho_{\rm TC}), (v_{e1},v_{e2}))=1$.
In the second one, ${\cal E}^{ZX}(\cdot)$ does not contain any string operators connecting $(v_{e1}, v_{e2})$,
and the above schematic discussion indicates the vanishing of the correlator for that case. 
In order to show this, let us consider the case that ${\cal E}^{ZX}(\cdot)$ contains the string $\tilde{W}^{ZX}_{12}$
but $Z_{\ell_0}X_{\ell_0+\delta}$, that is, the string $\tilde{W}^{ZX}_{12}$ gets separated into two parts.
(See Fig.~\ref{Fig5}(b).)
In this case, 
\begin{eqnarray}
&& C^{\rm II}({\cal E}^{ZX}(\rho_{\rm TC}), (v_{e1},v_{e2})) \nonumber \\
&& \hspace{1cm}\propto
\mbox{Tr} [{\cal E}^{ZX}(\rho_{\rm TC}) {\cal E}^{ZX}(Z_{\ell_0}X_{\ell_0+\vec{\delta}}\rho_{\rm TC}Z_{\ell_0}X_{\ell_0+\vec{\delta}})] =0,
\nonumber
\end{eqnarray}
as some of the toric code stabilizers anti-commute with $Z_{\ell_0}X_{\ell_0+\vec{\delta}}$.
More general cases can be considered similarly by replacing $Z_{\ell_0}X_{\ell_0+\vec{\delta}}$ with lacking string operators.

\section{Summery of aspect of symmetry and numerical observation}
Finally, let us summarize the 1-form symmetry properties of the model and their evolution through the stochastic decoherence, which we observed by the numerical methods. 
The system that we focus on consists of three parts: initial density matrix, decohered state and the decoherence channel $\mathcal{E}^{ZX}_\ell$. 
The initial state is the toric code $\rho_{\rm TC}$, and the decohered state denoted by $\rho_D$ is an ensemble of density matrices generated by the stochastic decoherence channels. 
Physical quantities are calculated by taking the average over states in the ensemble.
We find that the phase transition takes place, and here, we focus on decohered states for $r>r_c$.  
Then, each of the three parts holds the following property concerning the 1-form symmetry $W^{ZX}_{\gamma_c}$. 
\begin{table}[h]
\begin{tabular}{ |c||c|c| } 
\hline
    & strong & weak \\\hline
$\rho_{TC}$ & $\circ$ & $\circ$ \\
$\mathcal{E}^{ZX}_\ell$ & $\times$ & $\circ$  \\ 
$\rho_{D}$ & $\times$ & $\circ$  \\ 
\hline
\end{tabular}
\caption{Property with respect to 1-form symmetries $W^{ZX}_{\gamma_c}$ for toric code, decoherence channel and decohered state, respectively. 
For the states, $\rho_{TC}$ and $\rho_D$, some of the symmetries are SSB. See the table below.}
\end{table}

Details of the above observation are explained in Appendix B 1. 
All three objects are weak-symmetric with respect to the 1-form symmetry generated by $W^{ZX}_{\gamma_c}$.

With understanding the symmetry properties in Table I, we numerically calculated the following order and disorder parameters $C^{\rm I}(\rho,\gamma'_c)$ and $C^{{\rm II}}(\rho,(v_{e1},v_{e2}))$ for studying possible SSB of 1-form weak symmetry generated by $W^{ZX}_{\gamma_c}$ in the ensemble level of the stochastic decoherence system. 
By using these quantities, we obtained the results in Table II.
\begin{table}[h]
\begin{tabular}{ |c||c|c|c| } 
\hline
    &$C^{\rm I}(\rho,\gamma'_c)$ & $C^{{\rm II}}(\rho,(v_{e1},v_{e2}))$ & \mbox{order} \\\hline
$\rho_{\rm TC}$ & $\mathcal{O}(1)$ & $0$ & weak SSB \\
$\rho_{D}(r:{\mbox{large}})$ & $0$ & $\mathcal{O}(1)$ & weak symmetric \\
\hline
\end{tabular}
\caption{Orders for 1-form weak symmetries of $W^{ZX}_{\gamma_c}$.}
\end{table}
\\
Consequently, we numerically observed the transition between the states listed in Table II. 
The decoherence transition induced by $\mathcal{E}^{ZX}_\ell$ in the stochastic system is to be understood from the viewpoint of the SSB of
the 1-form weak symmetry $W^{ZX}_{\gamma_c}$.  
In conclusion, the transition from weak-symmetry SSB state (the TC state) to the decohered symmetry-restored state takes place.

\section{Discussion and conclusion}
In this letter, we gave a concrete demonstration of the emergence of the IMTO from the genuine topological state (the TC state) in the stochastic ZX-decoherence channel. 
In particular, we found the phase transition that is understood as the restoration of the 1-form weak symmetry and anyon proliferation, and verified the expectation that its phase transition is closely related to the 2D percolation by estimating the critical exponents.
Then, we comment that studying the relation to subsystem code~\cite{poulin2005,bacon2006,sohal2024,ellison2023,KI2023} is an interesting future direction. In addition, investigating the relation between critical exponents and the percolation picture in detail will be a future research topic and also
study of higher dimensional systems having similar setups to this work may be interesting.

The present work focused on a specific type of lattice model and discussed the specific types of the strong or weak symmetries. 
It is very interesting to consider an effective field theory in the continuum to describe this phase transition, because in the continuum there is no distinction between the two microscopic fermionic string operators. 
Subtleties regarding microscopic details of string operators in lattice models
can be further explored by considering decoherence of other lattice models such
as string-net models and Kitaev's honeycomb model.

\section*{ACKNOWLEDGMENTS}
This work is supported by JSPS KAKENHI: JP23K13026(Y.K.) and JP23KJ0360(T.O.). 
We acknowledge two anonymous referees for many insightful suggestions. In particular, one of them gave important suggestions and understandings about (I) the behavior of the ensemble average of $\mathbb{E}[C^{\rm I}(\rho_{s},\gamma'_c)]$ in Sec.V B, (II) the relationship between between the decoherence phenomenon and the 2D percolation in Sec.VI.


\section*{Appendix}

\subsection{Definition of strong and weak symmetry for density matrix}
We give a brief summery for the definition of the strong and weak symmetries for a density matrix.

In general, a density matrix (mixed state) can have two types of symmetries: strong and weak symmetries.  
The condition of the strong symmetry to a state $\rho$ is given as \cite{groot2022}
\begin{eqnarray}
U_g\rho=e^{i\theta}\rho,\nonumber
\end{eqnarray}
where $\rho$ is a state (pure or mixed) and $U_g$ is a symmetry operation of an element $g$ of a symmetry group $G$ and $\theta$ is a certain global phase factor. 

Next, the weak symmetry is characterized by
\begin{eqnarray}
U_{g}\rho U^\dagger_{g} =\rho.\nonumber
\end{eqnarray}
This condition is called the average or weak symmetry \cite{ma2024}, as the symmetry is satisfied in 
the ensemble in general. 
Note that if a state $\rho$ has a strong symmetry, the state is also weak symmetric with respect to that symmetry.

These notions are also applicable to quantum channels  $\mathcal{E}(\rho)$. 
Here, we consider the operator-sum representation of the channel \cite{Nielsen2011},
\begin{eqnarray}
\mathcal{E}(\rho)=\sum^{N-1}_{\ell=0}K_{\ell} \rho K^\dagger_{\ell},\nonumber
\end{eqnarray}
where $K_{\ell}$'s are Kraus operators satisfying $\sum^{N-1}_{\ell=0} K^\dagger_{\ell} K_{\ell}=I$. 
The channel $\mathcal{E}$ changes the input state $\rho$, and it includes non-unitary transformations such as decoherence and quantum measurements.

For the channel $\mathcal{E}$, the strong symmetry condition on the channel for a symmetry $G$ is represented as \cite{groot2022}
$$K_{\ell}U_g=e^{i\theta} U_g K_{\ell}$$
for $\forall \ell$ and $g \in G$, where $\theta$ is a single phase. 
On the other hand, weak symmetry condition on the channel for a symmetry $G$ is given as 
\begin{eqnarray}
U_g\biggl[\sum_{\ell}K_{\ell} \rho K^\dagger_{\ell}\biggr]U^\dagger_g=\mathcal{E}(\rho).\nonumber
\end{eqnarray}
All  Kraus operators $\{K_{\ell}\}$ do not necessarily commute with $U_g$.

\subsection{Symmetry aspects for the system}
We can introduce various types of 1-form symmetries 
by considering loop Pauli string operators \cite{wang2024,sohal2024}. 
In what follows, we discuss some kinds of symmetries regarded as 1-form weak or strong symmetries, which play an important role for the classification of topological order (TO), and then consider their SSB \cite{zhang2024} for the typical pure and mixed states, $\rho_{\rm TC}$ and $\rho_f$ in the main text.

In the toric code (TC) state, $\rho_{\rm TC}$, there are two 1-form strong symmetries, which are often represented by the Wilson and 't Hooft loop operators, and the anyon content is given as $\{1,e,m,f\}$, where $e(m)$ referees to electric (magnetic ) anyon, as well as their fusion $f=(e\times m)$.
In the IMTO state, $\rho_f$, on the other hand, the structure of the 1-form symmetry is slightly complicated as 
the state emerges through the decoherence by Kraus operators, some of which are non-commutative. 
As a result, loop operators of the 1-form symmetries are a product of composite operators as we clarify in this section.\\

\noindent{\bf 1. Contractible loop $W^{ZX}$ type of 1-form symmetry:} 
The first 1-form symmetry in $\rho_f$ is given by a product of Kraus operators along an arbitrary contractible loop on $v$-lattice $\gamma_c$;
$W^{ZX}_{\gamma_{c}}=\prod_{\ell \in \gamma_{c}}Z_{\ell}X_{\ell+\vec{\delta}}$.
It is verified that $W^{ZX}_{\gamma_{c}}$ anti-commutes with some of Kraus operators but commutes with all of the stabilizers.
The channel $\mathcal{E}^{ZX}$ satisfies the condition of 1-form weak symmetry \cite{groot2022} since for a local channel for any link $\ell$, 
$
\mathcal{E}^{ZX}_{\ell}(\rho)=\frac{1}{2}\rho+\frac{1}{2}Z_{\ell}X_{\ell+\vec{\delta}}\rho Z_{\ell}X_{\ell+\vec{\delta}},
$
$
W^{ZX}_{\gamma_{c}}\mathcal{E}^{ZX}_{\ell}(\rho)W^{ZX\dagger}_{\gamma_{c}}=\mathcal{E}^{ZX}_\ell(\rho)$  is satisfied.
As we stated in the above, the operators $W^{ZX}_{\gamma_{c}}$ commutes with the stabilizers  
$\{W_v=A_v B_{v+\vec{\delta}}\}$, and therefore the mixed state $\rho_f\propto \prod_v (1+W_v)/2$ is 1-form weak symmetric; 
$W^{ZX}_{\gamma_{c}} \rho_{f} W^{ZX\dagger}_{\gamma_{c}}=\rho_{f}$.
However, one should note that $W^{ZX}$ symmetry is \textit{not} a strong symmetry of $\rho_f$ as 
$W^{ZX}_{\gamma_{c}} \rho_{f} \neq \rho_{f}$.

On the other hand, it is easily verified that the TC state $\rho_{\rm TC}$ is also weak symmetric for 
$W^{ZX}_{\gamma_{c}}$,  i.e., $W^{ZX}_{\gamma_{c}} \rho_{\rm TC} W^{ZX\dagger}_{\gamma_{c}}=\rho_{\rm TC}$, which directly 
comes from the fact that the state $\rho_{\rm TC}$ satisfies the strong symmetry condition; $W^{ZX}_{\gamma_{c}} \rho_{\rm TC}=\rho_{\rm TC}$.
Existence of a strong symmetry in a state guarantees that the state has the corresponding weak symmetry.
Both $\rho_{\rm TC}$ and $\rho_f$ are weak-symmetric
under the 1-form symmetry $W^{ZX}_{\gamma_{c}}$.

By the standard manipulation, we can introduce an anyon corresponding to the
1-form symmetry $W^{ZX}$ by considering an operator  $W^{ZX}_{\cal C}$, where ${\cal C}$ is an open string and a pair of
anyons emerge on endpoints of ${\cal C}$. 
We call this anyon $f_w$-anyon, which plays an important role for the discussion on the proliferation of anyon 
under decoherence as explained in the main text.

By following Ref.~\cite{zhang2024},  
one can study whether the $W^{ZX}$ symmetry is spontaneously broken or not. 
To characterize the SSB of the 1-form weak symmetry of $W^{ZX}$,  order and disorder parameters are to be used; the order parameter is given by
$$
C^{\rm I}(\rho,\gamma'_c) \equiv \frac{\Tr[\rho W^{X(Z)}_{\gamma'_{c}} \rho]}{\Tr[\rho^2]},
$$ 
where $W^{X(Z)}_{\gamma'_{c}}=\prod_{\ell \in \gamma'_{c}}X_{\ell}(Z_{\ell})$ is a 't Hooft (Wilson) loop braiding nontrivially with $W^{ZX}_{\gamma_c}$. 

On the other hand for the disorder parameter, the following R\'enyi-2 correlator is used,
$$
C^{{\rm II}}(\rho,(v_{e1},v_{e2})) \equiv \frac{\Tr[\rho W^{ZX}(v_{e1},v_{e2}) \rho W^{ZX}(v_{e1},v_{e2})]}{\Tr[\rho^2]},
$$
where the operator
$W^{ZX}(v_{e1},v_{e2})$ is defined similarly to the symmetry operator $W^{ZX}_{\gamma_c}$ supported along an open string with endpoints
$(v_{e1},v_{e2})$ as shown in Fig.~\ref{Fig1}(b) in the main text.
As explained in the above, a pair of $f_w$-anyons are produced at $v_{e1}$ and $v_{e2}$.
As $W^{ZX}(v_{e1},v_{e2})$ is commutative with all the stabilizers, $f_w$-anyon is a non-detectable anyon.

If the state $\rho$ exhibits the1-form $Z_2$ weak SSB (for $W^{ZX}$), then 
$$
C^{\rm I}(\rho,\gamma'_c)=\mathcal{O}(1),\:\: 
C^{{\rm II}}(\rho,(v_{e1},v_{e2})) =0.
$$
If not, 
$$
C^{\rm I}(\rho,\gamma'_c)=0,\:\: 
C^{{\rm II}}(\rho,(v_{e1},v_{e2})) = \mathcal{O}(1).
$$

In the main text, to characterize the IMTO numerically, we study the order and disorder parameters,
$C^{\rm I}(\rho,\gamma'_c)$ and  $C^{{\rm II}}(\rho,(v_{e1},v_{e2}))$. 
In particular, in the stochastic ZX-decoherence channel, we observe the averaged susceptibility obtained from data of many samples.\\

\noindent{\bf 2. Contractible loop $W^{XZ}$ type of 1-form symmetry:}
By following Refs.~\cite{wang2024,sohal2024}, 
another important 1-form symmetry is introduced, which is commutative with all the Kraus operators
of the decoherence and also $W^{ZX}_{\gamma_{c}}$.
The symmetry operator is given by  
$W^{XZ}_{\gamma_{c}}=\prod_{\ell \in \gamma_{c}}X_{\ell}Z_{\ell+\vec{\delta}}$,  
where again, $\gamma_{c}$ is a contractible closed loop on the $q$-lattice.
It is easily verified that the channel $\mathcal{E}^{ZX}_\ell$ is strong symmetric for $W^{XZ}$ \cite{groot2022} due to $[W^{XZ}_{\gamma_{c}},Z_\ell X_{\ell+\vec{\delta}}]=0$ for $\forall \ell$.
The maximal decohered ZX state $\rho_f$  
exhibits the 1-form strong symmetry for $W^{XZ}_{\gamma_{c}}$, $W^{XZ}_{\gamma_{c}} \rho_{f}=\rho_{f}  
$ since $W^{XZ}_{\gamma_{c}}$ is given by a product of 
the stabilizers $W_v$ \cite{wang2024,sohal2024}. 
The TC state $\rho_{\rm TC}$ is also strong symmetric, i.e., $W^{XZ}_{\gamma_{c}} \rho_{\rm TC}=\rho_{\rm TC}$.

By following Ref.~\cite{zhang2024}, there are measures to investigate whether the 1-form strong symmetry of $W^{XZ}_{\gamma_{c}}$ is spontaneously broken or not, 
that is, one can introduce the strong-SSB order and disorder parameters for $W^{XZ}_{\gamma_{c}}$ symmetry. 
[Here ``strong-SSB'' means that we focus on the SSB order/disorder parameters for 1-form symmetry in strong symmetry sense.
These are different from those in weak symmetry sense.
See later discussion.]
The order parameter is given by 
$$
O_2(\rho) \equiv \frac{\Tr[W^{X(Z)}_{\gamma_{c}} \rho W^{X(Z)}_{\gamma_{c}}\rho ]}{\Tr[\rho^2]},
$$ 
where $W^{X(Z)}_{\gamma_{c}}=\prod_{\ell \in \gamma_{c}}X_{\ell}(Z_{\ell})$ is again a 't Hooft (Wilson) loop for a contractible loop $\gamma_{c}$, braiding nontrivially with $W^{XZ}_{\gamma_c}$.
On the other hand, the strong-SSB disorder parameter is given as
$$
D_1(\rho) \equiv \frac{\Tr[\rho W^{XZ}(q_{e1},q_{e2}) \rho ]}{\Tr[\rho^2]},
$$
where 
$W^{XZ}(q_{e1},q_{e2})$ is an operator obtained from $W^{XZ}_{\gamma_{loop}}$ by restricting its support to an open string with endpoints labeled by $q_{e1}$ and $q_{e2}$ on the $q$-lattice. 
Similarly to the $W^{ZX}$ 1-form symmetry, the above operator for an open string 
$ W^{XZ}(q_{e1},q_{e2})$ accompanies a pair of anyons at the endpoints $q_{e1},q_{e2}$, which we call $f_s$-anyon in the main text.
As $W^{XZ}(q_{e1},q_{e2})$ anti-commutes with the stabilizers at $q_{e1}$ and $q_{e2}$, $f_s$-anyon is a detectable anyon.

For the typical states $\rho_{\rm TC}$ and $\rho_f$, we can observe how the above order and disorder parameters 
for the 1-form strong XZ-symmetry behave. 
The maximal ZX-decohered state $\rho_{f}$ is a SSB state of the 1-form strong symmetry since 
the 't Hooft and Wilson loops commute with all the stabilizers and the string operator $W^{XZ}(q_{e1},q_{e2})$ cannot be given by a product of the stabilizers indicating $O_2(\rho_f)=1,\:\: D_1(\rho_f)=0$.  
It is similarly verified that the TC state $\rho_{\rm TC}$ is also 1-form strong SSB state since 
$
O_2(\rho_{\rm TC})=1,\:\: 
D_1(\rho_{\rm TC})=0$. 

Furthermore, one can study the properties of $W^{XZ}_{\gamma_c}$ as a 1-form weak symmetry.
The order and disorder parameters are given as before,
$$
O_1(\rho)=\frac{\Tr[\rho W^{X(Z)}_{\gamma_{c}} \rho ]}{\Tr[\rho^2]},
$$
$$
D_2(\rho)=\frac{\Tr[W^{XZ}(q_{e1},q_{e2})\rho W^{XZ}(q_{e1},q_{e2}) \rho ]}{\Tr[\rho^2]}.
$$

Let us observe the above quantities for the states, $\rho_{\rm TC}$ and $\rho_{f}$. 
For $\rho_f$, 
$$
O_1(\rho_f)=0, \:\:\: D_2(\rho_f)=0, 
$$
since $W^{X(Z)}_{\gamma_{c}}$ is not an element of the stabilizer group of the state $\rho_f$ [$O_1(\rho_f)=0$], and $W^{XZ}(q_{e1},q_{e2})$ 
anti-commutes with the stabilizes $W_v$'s at the endpoints [$D_2(\rho_f)=0$].
As both the order and disorder parameters are vanishing, we cannot judge if 1-form weak SSB for 
$W^{XZ}_{\gamma_{c}}$ symmetry takes place or not.
This result is commented on later on.

On the other hand for the state $\rho_{\rm TC}$, 
$$
O_1(\rho_{\rm TC})=1,\:\:\: D_2(\rho_{\rm TC})=0,
$$
since $W^{X(Z)}_{\gamma_{c}}$ is a product of $A_v$'s or $B_p$'s, and the endpoint of $W^{XZ}(q_{e1},q_{e2})$ anti-commutes with some $A_v$ and $B_p$.
Thus, the state $\rho_{\rm TC}$ is 1-form weak SSB for $W^{XZ}_{\gamma_{c}}$. In other words, the TC is a strong-to-trivial SSB state.\\

\noindent{\bf 3. Comments on $XZ$-type 1-form symmetry with non-contractible loop}: 
We can also introduce $XZ$-type 1-form symmetry with non-contractible loop. 
The generator is given by
$$
W^{XZ}_{\gamma_{nc}}=\prod_{\ell \in \gamma_{nc}}X_{\ell}Z_{\ell+\vec{\delta}},
$$
where $\gamma_{nc}$ is a non-contractible loop on $q$-lattice. 
As this operator commutes with every Kraus operator, 
it remains a 1-form strong symmetry under 
the decoherence channel $\mathcal{E}^{ZX}_\ell$.
However, the ability of the logical operator is lost as its quantum conjugate operator such as 
the Wilson and 't Hooft non-contractible loop operators disappear from the stabilizer group
under the channel $\mathcal{E}^{ZX}_\ell$ even if they are contained in the stabilizer group of the initial pure state.
In other words, a transition from the genuine topologically ordered (TO) state to a state with classical memory takes place. 
More precisely, the proposed disorder parameters do not work for the non-contractible
1-form symmetry, because a finite length string is used in $D_1$ and $D_2$.
\\

\noindent{\bf 4. Lacking of disorder parameter for $XZ$-type 1-form symmetry with detectable anyon}:
When an open string operator accompanies detectable anyon at endpoints ($f_s$-anyon in the current case) as in the 1-form strong $W^{XZ}$ symmetry considered in the above, the disorder parameter $D_2 \sim \langle  U_g (q_{e1},q_{e2}) \rho U^\dagger_g (q_{e1},q_{e2}) \rho \rangle$ is automatically zero because anyon is non-commutative with stabilizers at endpoints.
[Its non-commutativity with stabilizers is the definition of the detectable anyon.]
The order parameter of weak symmetry, $O_1$, may also be zero, and in this case, it is unable to judge how
the symmetry is realized in that state.
In the current case, $O_1$ is zero for all possible $U_h$ for the state $\rho_f$.
This result indicates that order and disorder parameters for 1-form symmetry have {\it not} been perfectly prepared yet, and further study on it is desired.


\subsection{Numerical scheme}
\noindent{\bf 1. Update method for ZX decoherence in the stabilizer formalism}: 
Some of decoherence channel can be implemented in the efficient stabilizer algorithm \cite{weinstein2022}.
One example is a projective measurement without monitoring. 
We consider a local maximal decoherence corresponding to measurements with a local measure operator $\hat{m}_{i}$ without monitoring (recording) the outcomes, where the label $i$ refers to a relevant portion of the system.  
Here, we assume that $\hat{m}_{i}$ is an element of Pauli group with $+1$ factor, and its measurement outcome denoted by $\beta_i$ takes $\beta_i=\pm 1$. 
The channel is then given by 
\begin{eqnarray}
\mathcal{E}^{\hat{m}}_i[\rho]=\sum_{\beta_i=\pm}P^{m_i}_{\beta_i}\rho P^{m_i\dagger}_{\beta_i}=\frac{1}{2}\rho+\frac{1}{2}\hat{m}_i\rho \hat{m}_{i},\nonumber
\end{eqnarray}
where $P^{m_i}_{\beta_i}$ is a projection operator for $\hat{m}_i$ with the outcome $\beta_i$ given by $P^{m_i}_{\beta_i}=\frac{1+\beta_i \hat{m}_i}{2}$.
If we apply the channel to an entire system with $L$ degrees of freedom, the total channel is represented as
\begin{eqnarray}
\mathcal{E}^{\hat{m}}[\rho]= \prod^{L-1}_{i=0} \mathcal{E}^{\hat{m}}_i[\rho].\nonumber
\end{eqnarray}

Let us explain how the local channel $\mathcal{E}^{\hat{m}}_i$ acts to a mixed state in the stabilizer formalism.  
Here, we consider a density matrix represented by stabilizer generators $\{g_{n}\}$,
\begin{eqnarray}
\rho=\frac{1}{2^{L-k}}\prod^{k-1}_{n=0}\frac{1+g_{n}}{2},
\label{dens_stab}
\end{eqnarray}
where $L$ is the total number of qubit in a system and $k$ is the total number of independent stabilizer generators (generally, $k \le L$). 
According to Ref.~\cite{weinstein2022}, the introduction of the local decoherence channel $\mathcal{E}^{\hat{m}}_i$ is efficiently implemented in the stabilizer algorithm. 
When one applies $\mathcal{E}^{\hat{m}}_i$ to $\rho$, then the state $\rho$ becomes
\begin{eqnarray}
\mathcal{E}^{\hat{m}}_i[\rho]&=&\sum_{\beta_i=\pm}P^{m_i}_{\beta_i}\rho P^{m_i\dagger}_{\beta_i}\nonumber\\
&=&\sum_{\beta_i=\pm}P^{m_i}_{\beta_i} \biggl[\frac{1}{2^{L-k}}\prod^{k-1}_{n=0}\frac{1+\tilde{g}_{n}}{2}\biggr]P^{m_i\dagger}_{\beta_i}\nonumber\\
&=&\sum_{\beta_i=\pm}P^{m_i}_{\beta_i} \biggl[\frac{1+\tilde{g}_{0}}{2}\biggr]P^{m_i\dagger}_{\beta_i} \biggl[\frac{1}{2^{L-k}}\prod^{k-1}_{n=1}\frac{1+\tilde{g}_{n}}{2}\biggr]\nonumber\\
&=&\frac{1}{2^{L-k+1}}\prod^{k-1}_{n=1}\frac{1+\tilde{g}_{n}}{2},\nonumber
\end{eqnarray}
where on the second line we have performed a standard transformation (corresponding to the change of the representation of the stabilizer generator) \cite{Nielsen2011} to obtain other representation of  $\{\tilde{g}_n\}$, in which at most one stabilizer generator labeled by $\tilde{g}_{0}$ is anti-commutative with $\hat{m}_i$. 
Thus, from the last line of the above equations, the application of the local channel $\mathcal{E}^{\hat{m}}_i$ eliminates a single stabilizer generator from the set of stabilizer generators, leading to the enhancement of the mixing of the state. 
This procedure is directly implemented in numerical simulations.\\

\noindent{\bf 2. Numerical calculation of negativity}: 
We briefly explain how to calculate the negativity $N_A=\log_2|\rho^{\Gamma_A}|_1$ defined in the main text. 
The original method calculating the negativity and its exact derivation are described in Refs.~\cite{sang2021,shi2021}. 
As explained in Refs.~\cite{sang2021,shi2021,sharma2022,KOI2023_1}, the negativity $N_A$ is numerically obtained from a $\mathbb{F}_2$ matrix $J$ 
\begin{eqnarray}
N_{A}=\frac{1}{2}\mathrm{rank}J,\nonumber
\end{eqnarray}
where $J$ is a $m \times m$ matrix, $m$ is the total number of the stabilizer generator for a state $\rho$ in $L$ total qubit system and here $m \le L$.
The matrix $J$ is constructed from $m$-stabilizer generators $g_{n}$ ($n=0,\cdots, m-1$) of the state $\rho$. Here, let us use the binary representation for $g_{n}$ \cite{Nielsen2011}. 
Then, we truncate the binary representation vectors of each stabilizer generator as 
\begin{eqnarray}
g_{n} \longrightarrow g^A_{n}=(g^{n,x}_0,\cdots, g^{n,x}_{k} | g^{n,z}_0,\cdots, g^{n,z}_{k}),\nonumber
\end{eqnarray} 
where $g^{n,x(z)}_{p}=0$ or $1$ and the binary components with the qubit label of the subsystem $A$ (here labeled by $0,\cdots, k$) remain.  
Finally, by using the $m$ truncated stabilizer generators $g^A_{n}$, we construct the matrix $J$ given by 
\begin{eqnarray}
(J)_{n,n'}=
\begin{cases}
1 & \mbox{if}\:\: \{g^{A}_{n},g^{A}_{n'}\}=0\\
0 & \mbox{if}\:\: [ g^{A}_{n},g^{A}_{n'}]=0
\end{cases}
,\nonumber
\end{eqnarray}
where $\{g^{A}_{n},g^{A}_{n'}\}=0$ means that the truncated stabilizer generators $g^{A}_{n}$ and $g^{A}_{n'}$ are anti-commuting,
and similarly $[g^{A}_{n},g^{A}_{n'}]=0$ means that the truncated stabilizer generators $g^{A}_{n}$ and $g^{A}_{n'}$ are commuting.
By this manipulation, we obtain the binary $m \times m$ $\mathbb{F}_2$ matrix $J$. Finally, the calculation of the rank of $J$ gives the value of negativity $N_A$.\\

\noindent{\bf 3. Calculation method of $C^{\rm I}$:}
We explain the numerical method for calculating the following quantity introduced in the main text, 
$$
C^{\rm I}(\rho,\gamma'_c) = \frac{\Tr[\rho W^{Z}_{\gamma'_c} \rho ]}{\Tr[\rho^2]}.
$$
If the operator $W^{Z}_{\gamma'_c}$ is a stabilizer obtained by a product of the stabilizer generators $g_n$ of the state $\rho$, then $C^{\rm I}(\rho,\gamma'_c)=1$ since $W^{Z}_{\gamma'_c}\rho=\rho$.
On the other hand, if the operator $W^{Z}_{\gamma'_c}$ is linearly
independent from all stabilizer generators of $\rho$, 
then $C^{\rm I}(\rho,\gamma'_c)=0$. 
Further, if the operator $W^{Z}_{\gamma'_c}$ is anti-commutative to at least one stabilizer generator of $\rho$, then $C^{\rm I}(\rho,\gamma'_c)=0$. These three cases are numerically judged by using the check matrix and the basic transformations of the set of stabilizer generators \cite{Nielsen2011}. Thus, we can numerically calculate the value of $C^{\rm I}(\rho,\gamma'_c)$.\\
\begin{figure}[t]
\begin{center} 
\vspace{0.5cm}
\includegraphics[width=7cm]{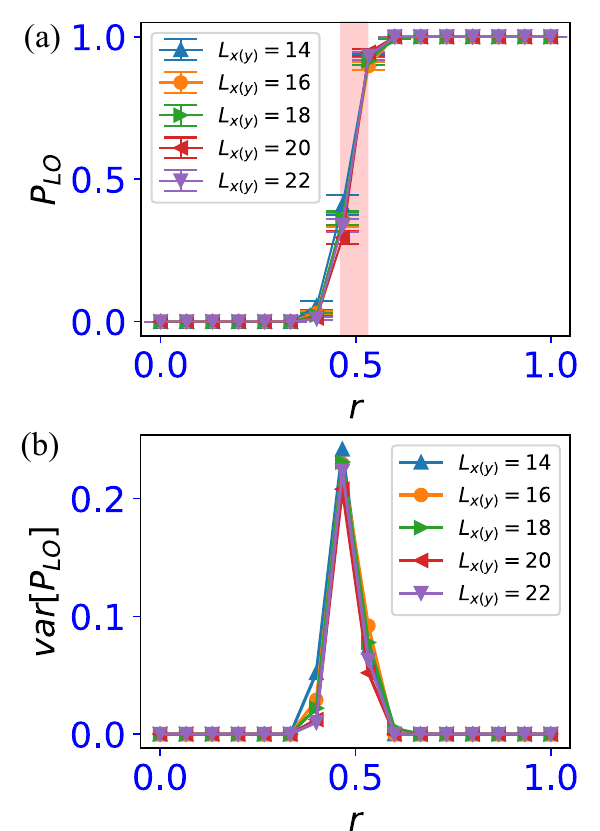}  
\end{center} 
\caption{(a) Annihilation probability of the logical operator of the TC state, $W^{X}_{\gamma^{x(y)}}$. (b) Variance of the annihilation probability obtained by averaging over samples of the trajectory.}
\label{FigPLO}
\end{figure}

\noindent{\bf 4. Calculation method of R\'enyi-2 correlator:}
The R\'enyi-2 correlator such as $C^{{\rm II}}(\rho,(v_{e1},v_{e2}))$ in the main text can be calculated in the stabilizer formalism. 
We introduce an operator $O_i$. 
We assume that the operator $O_i$ belongs to the Pauli group including identity with the factor $+1$ and a density matrix $\rho$ is given by Eq.~(\ref{dens_stab}).  
Since each stabilizer generator $g_{n}$ commutes or anti-commutes with the operator $O_i$, the R\'enyi-2 correlator for the density matrix $\rho$ is given as 
\begin{eqnarray}
&&C^{\rm II}_{\rm O_iO_j}[\rho]=\frac{\Tr[O_iO_j\rho O_jO_i\rho]}{\Tr[\rho^2]} \nonumber,
\end{eqnarray}
and the numerator and denominator are calculated as 
\begin{eqnarray}
&&\Tr[O_i O_j\rho O_j O_i\rho]=\frac{1}{2^{L}}\biggl[\prod^{k-1}_{n=0}(1+\alpha^n)\biggl],\nonumber\\
&&\Tr[\rho^2]=\frac{1}{2^{L-k}},\nonumber
\end{eqnarray}
with the factor $\alpha^n=\pm$ for $[O_iO_j,g_n]_{\pm}=0$ (where $[\cdot]_{\pm}$ is commutative or anti-commutative bracket) and we have used $O_iO_j(1+g_{n})O_jO_i=(1+\alpha^{n}g_{n})$. 
Thus, we only extract the (anti)-commutativity between $O_iO_j$ and each stabilizer generator to obtain the R\'enyi-2 correlator.\\

\subsection{Annihilation probability of logical operator of toric code}
As a supplemental observation, we numerically observe the fate of the logical operator of the TC $W^{X}_{\gamma_{x(y)}}$ by applying the stochastic ZX-decoherence channel discussed in the main text. 

We prepare the pure unique TC state with the logical operators given by the set of the stabilizer generator, $S'_{\rm int}=S_{\rm int}+S_{X-{\rm lg}}$. 
For a sparse ZX-decoherence corresponding to a small probability $r$, the logical operators can survive
by deforming the loop $\gamma_{x(y)}$ due to applying star operator $A_v$'s.
For a dense ZX-decoherence corresponding to a large $r$, the logical operator is strongly deformed and beyond a threshold, it is to be swept away from the set of the stabilizer generator, indicating 
vanishing of the TC phase. 
We numerically trace the existence of the logical operators under the numerical update in the efficient stabilizer update algorithm and estimate the annihilation probability of the logical operators denoted by $P_{LO}$ as varying the probability $r$ 
(A similar calculation for a pure state case has already been demonstrated in \cite{botzung2023}). 
Figure~\ref{FigPLO}(a) is the behavior of $P_{LO}$ obtained by averaging over trajectories. 
$P_{LO}$ suddenly increases around $r=0.5$ consistent with the numerical results shown in the main text, that is, the topological order of the TC state is swept indicating the change of the mixed state in this stochastic circuit. Finally, the variance of $P_{LO}$ obtained by the fluctuation of the trajectory samples is shown in Fig.~\ref{FigPLO}(b). 
The peak is observed around $r=0.5$, indicating the mixed state phase transition in the level of the averaged trajectory picture. 

\bibliography{ref}

\begin{thebibliography}{61}
\bibitem [{com()}]{come_TC_SSB}%
  \BibitemOpen
  \href@noop {} {\ }\bibinfo {note} {Note that the original state $\rho_{TC}$ is 1-form strong symmetric for $W^{ZX}_{\gamma'_c}$ as $W^{ZX}_{\gamma'_c}\rho_{TC}=\rho_{TC}$. But, we need to include the channel decoherence dynamics $\mathcal{E}^{ZX}$, which is not strong symmetric. Thus, we regard that $\rho_{TC}$ is also weak symmetric.}\BibitemShut {Stop}%
\end{thebibliography}

\end{document}